\documentclass[prb,amsmath,amssymb]{revtex4}

\usepackage{graphicx}
\usepackage{dcolumn}
\usepackage{bm}

\begin{document}


\title{Ab initio calculation of anisotropic interfacial excess free energies}

\author{A. van de Walle}
\affiliation{Brown University}
\email{avdw@alum.mit.edu}
\author{C. Balaji Gopal}
\affiliation{Caltech}
\author{S. Demers}
\affiliation{Caltech}
\author{Q. Hong}
\affiliation{Brown University}
\author{A. Kowalski}
\affiliation{Google}
\author{L. Miljacic}
\affiliation{Brown University}
\author{G. Pomrehn}
\affiliation{Caltech}
\author{P. Tiwary}
\affiliation{ETHZ}

\date{\today}

\begin{abstract}
We describe\ a simple method to determine,\ from ab initio calculations, the
complete orientation-dependence of interfacial free energies in solid-state
crystalline systems. We illustrate the method with an application to
precipitates in the Al-Ti alloy system.\ The method combines the cluster
expansion formalism in its most general form (to model the system's
energetics) with the inversion of the well-known Wulff construction (to
recover interfacial energies from equilibrium precipitate shapes). Although
the inverse Wulff construction only provides the relative magnitude of the
various interfacial free energies, absolute free energies can be recovered
from a calculation of a single, conveniently chosen, planar\ interface.\ The
method is able to account for essentially all sources of entropy (arising
from phonons, bulk point defects, as well as interface roughness) and is
thus able to transparently handle both atomically smooth and rough
interfaces. The approach expresses the
resulting orientation-dependence of the interfacial properties using
symmetry-adapted bases for general orientation-dependent quantities. As a
by-product, this paper thus provides a simple and general method to generate
such basis functions, which prove useful in a variety of other applications,
for instance to represent the anisotropy of the so-called constituent strain
elastic energy.
\end{abstract}

\maketitle

\section{Introduction}

Interfacial free energies play an important role in the design of
engineering materials, as they represent a fundamental determinant of
microstructure \cite{asta:modinterf,wolfyip:interf}. However, the
experimental measurement of interfacial free energies can represent a
difficult challenge \cite{lifshitz:nial,voorhees:nial,ardell:is,ardell:jms}
and computational approaches have thus proven useful to provide
complementary corroborating estimates
\cite{asta:modinterf,mishin:nialinterf,seidman:nial,ozolins:nialnat,asta:agal,asta:alliinterf,woodward:nial,muller:precip}.

The two main challenges faced when determining interfacial properties via
computational means are that (i) the interface structure may take
considerable time to equilibrate and that (ii) at finite temperature,
defects are present in thermodynamic equilibrium, a situation which
typically requires sampling of numerous possible microscopic states. This
situation prompts for the use of computationally very efficient energy
models. In the case of coherent interfaces, the cluster expansion formalism\ 
\cite{sanchez:cexp} has been proven a very effective approach to the
simulation of interfaces
\cite{ozolins:nialnat,asta:agal,asta:alliinterf,woodward:nial,muller:precip,avdw:apbtial,asta:interfalsc,zarkevich:agal,asta:fppheq,sluiter:alliinterf,avdw:smceo2}.
The cluster expansion provides a very compact and systematically
improvable representation of the configurational-dependence of an alloy's
energy and the process of constructing such a cluster expansion (with a
given accuracy)\ from ab initio data is well-established \cite{avdw:maps}.

While it is common to proceed by studying planar interfaces one direction at
a time in a supercell geometry, there are compelling reasons to proceed by
considering all interface directions in a single large-scale simulation.
This can be useful to directly determine equilibrium precipitate shapes
(e.g. \cite{muller:precip}). More generally, such a global approach proves
most useful for the determination of interface free energies if (i) the
thermodynamically stable directions are not known a priori or if\ there may
be a continuum of stable directions when temperature is sufficiently high
and the interfaces roughen or (iii) if the determination of the complete
orientation dependence of the interfacial free energy is needed.\ The latter
is especially useful to provide input to continuum-type simulations, such as
phase field \cite{moelans:phasefield,chen:aiphfi,chen:aiphfil}\ or finite
element \cite{raabe:finelem}\ modeling.

This paper describes a simple method to determine, from ab initio
calculations, the complete orientation-dependence of interface free energies
in solid-state crystalline systems.\ The method combines the cluster
expansion formalism in its most general form (to model the system's
energetics) with the inversion of the well-known Wulff construction \cite%
{wulff:wulff,laue:wulff,greene:iwulff} to recover interfacial energies from
equilibrium precipitate shapes. Although the inverse Wulff construction only
provides the relative magnitude of the various interfacial free energies,
absolute free energies can be recovered from a calculation of a single,
conveniently chosen, planar\ interface (although not necessarily flat at the
atomic level). The method is able to account for essentially all sources of
entropy (arising from phonons, bulk point defects, as well as interface
roughness) and is thus able to transparently handle both atomically smooth
and rough interfaces.\ To address the issue that some interface directions
do not appear on the Wulff shape when the facetting occurs, we show how the
interfacial free energy surface can be naturally extended into the
\textquotedblleft masked\textquotedblright\ regions of the Wulff plot in a
way that (i) preserves the predicted equilibrium shape and (ii) has a
natural geometric interpretation. The approach finally expresses the
resulting orientation-dependence of the interfacial properties using
symmetry-adapted bases for general orientation-dependent quantities. As a
by-product, this paper thus provides a simple and general method to generate
such basis functions, which prove useful in a variety of other applications,
for instance to represent the so-called constituent strain elastic energy of
superlattice structures in the long wavelength limit \cite%
{laks:recip,ozolins:elas}.\ The proposed method can be easily implemented
using generic linear algebra operations without having to consider many
different subcases that depend on the point group symmetry considered. The
method exploits a direct correspondence between spherical harmonics and
polynomial functions of a unit vector expressed in tensor notation. The
methods proposed herein have been implemented within the\ Alloy Theoretic
Automated Toolkit (ATAT) \cite{avdw:maps,avdw:atat,avdw:atat2,avdw:atatcode}.

As an example to demonstrate the method, we select the Ti-Al alloy system
because it exhibits a number of interesting and challenging features. Al$_{3}$%
Ti precipitates in an Al-rich fcc host exhibiting a D0$_{23}$ structure which
has a relatively low symmetry (thus increasing the number of distinct
facets). We also find that such precipitates exhibit a mixture of atomically
smooth and rough facets at all but the lowest temperatures.
Moreover, previous work
has shown that vibrational contributions to the free energy in this
alloy are large  \cite{avdw:jomatat,avdw:amape,liu:cdtfc} and must
therefore be included in the assessment of interfacial
boundary energies.

\section{Methods}

\subsection{Coarse Graining Cluster Expansion}

The system considered here consists of coherent precipitates within a host
phase with a known lattice type (fcc). For efficiency reasons, we select a
form of energy model especially adapted to this situation. The approach to
the determination of interface free energies is not tied to this specific
energy model, however. We rely on the cluster expansion formalism \cite%
{sanchez:cexp,fontaine:clusapp,zunger:NATO,ducastelle:book}, which
represents the energy $E$ of a crystalline alloy with a computationally
efficient Hamiltonian taking the form of a polynomial in terms of occupation
variables $\sigma _{i}=\pm 1$ indicating the type of atom\ residing on each
lattice site $i$:%
\begin{equation}
E=\sum_{i\not=j}J_{ij}\sigma _{i}\sigma
_{j}+\sum_{i\not=j\not=k}J_{ijk}\sigma _{i}\sigma _{j}\sigma _{k}+\ldots 
\label{eqclusexp}
\end{equation}%
\ The unknown coefficients, $J_{\cdots }$, of this polynomial are called
Effective Cluster Interactions (ECI) and are fit to a database of \emph{ab
initio} structural energies (obtained from Density Functional Theory total
energy calculations). It has been formally shown that an infinite series of
this form can represent any configuration-dependence of the energy \cite%
{sanchez:cexp}. Moreover, formal methods have been developed to determine
the number of terms and the database size needed to achieve a given
precision \cite{avdw:maps}. The ECI can typically be determined from a
reasonably-sized database of ab initio calculations involving
small-unit-cell atomic arrangements. These ECI can then be used in a
large-scale equilibrium Lattice-Gas Monte Carlo simulations of the coherent
interface, without necessitating repeated large-scale ab initio calculations
for each atomic configuration visited in a thermodynamic equilibrium. These
tasks were carried out with the help of the Alloy Theoretic Automated
Toolkit (ATAT) \cite{avdw:maps,avdw:atat,avdw:atat2,avdw:atatcode}.

The effect of lattice vibrations can be included within this framework, via
a coarse-graining technique \cite{avdw:vibrev}. The idea is to replace, in
Equation (\ref{eqclusexp}),\ the energy $E$ by the phonon contribution to
the free energy for a given configuration $\sigma $. The resulting effective
interactions $J_{\cdots }$ now become temperature-dependent, but the
formalism otherwise remains the same. As full lattice dynamics calculations
can be computationally expensive to repeat for numerous configurations $%
\sigma $, we rely on the \textquotedblleft bond stiffness vs bond
length\textquotedblright\ approach \cite{avdw:vibrev,avdw:pd3v}. In this
approach, the effective springs connecting nearby atoms are calculated for a
few structures only for a range of lattice parameters. The data thus
generated is used to determine the relationship between the stiffness of an
effective spring and the distance between the corresponding pair of atoms.
Once this is known, the relaxed atomic position for all remaining structures
(obtained from the ab initio calculations) 
can be used to predict spring constants without necessitating additional ab
initio lattice dynamics calculations.

The above cluster expansion deliberately does not include long-range elastic
effects, because we wish to determine the chemical and local relaxations
contribution of the interface,
and not the strain energy associated with
 deforming the host and
the precipitate due to their coherent coexistence. Had we included so-called
constituent strain effects \cite{laks:recip,ozolins:elas} in (\ref%
{eqclusexp}), we would have had to later subtract elastic energies via a
continuum elasticity-type analysis. One should however be cautious not to
interpret the equilibrium precipitate shape found in our simulations as the
actual equilibrium precipitate shape (which is affected by elastic effects).
However, the equilibrium precipitate shape we obtain is the relevant shape
for the purpose of determining the interfacial excess free energy,
independently of long-range elastic effects. For simplicity of the
presentation we nevertheless use the phrase \textquotedblleft equilibrium
precipitate shape\textquotedblright\ throughout, with the understanding that
elastic effects are omitted, as they should for a purely interfacial
analysis. It should be observed that, in the limit of small precipitates,
interfacial effects (scaling as $r^{2}$ for precipitates of length scale $r$%
) dominate over elastic effects (scaling as $r^{3}$), so our equilibrium
precipitate shapes should nevertheless be representative of actual
precipitate shapes in this $r\rightarrow 0$ limit.

\subsection{Ab initio calculations}

All ab initio calculations were performed with the VASP code \cite%
{kresse:vasp1,kresse:vasp2} implementing the Projector Augmented Wave (PAW)
method \cite{kresse:paw}. The PBE exchange-correlation functional \cite%
{pbe:pbe} was used. The kinetic energy cutoff was set to $300$ eV
(corresponding to VASP's\ \textquotedblleft high\textquotedblright\
precision setting). The k-point mesh was set via the algorithm described in 
\cite{avdw:maps} to ensure a density of at least $8000$ k-points per (atom)$%
^{-1}$ for all superstructures considered. For the large-supercell phonon
calculations, these settings were reduced $240$ eV (corresponding to VASP's\
\textquotedblleft medium\textquotedblright\ precision setting) and  $4000$
k-points per (atom)$^{-1}$, respectively.

\subsection{Inverse Wulff construction}

\begin{figure}
\centerline{\includegraphics{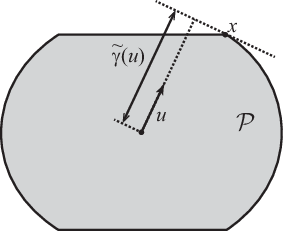}}
\caption{\label{iwulfffig}Graphical representation of the inverse Wulff construction defined by Equation (\ref{eqcgamma}).}
\end{figure}

The Wulff construction \cite{wulff:wulff,laue:wulff} is a well-known procedure to calculate the
equilibrium precipitate shape from the directional-dependence of the
interfacial excess free energy. This procedure can be readily inverted (see, e.g. \cite{greene:iwulff}) to
yield the interfacial excess free energy from the knowledge of the
equilibrium precipitate shape. Let $\gamma \left( u\right) $ denote the
interface free energy with an orientation defined by some unit vector $u$.
Let $\mathcal{P}$ denote a set in three-dimensional space representing the
shape of a precipitate in equilibrium. The inverse Wulff construction $\tilde{\gamma}\left(u\right)$
is illustrated in Figure \ref{iwulfffig} and is given by
\begin{equation}
\tilde{\gamma} \left( u\right) =\max_{x\in \mathcal{P}}u\cdot x.  \label{eqcgamma}
\end{equation}%
It follows directly from the Wulff construction that, for any facet $u$ part of the equilibrium precipitate
shape,
\begin{equation}
\gamma \left( u\right)=c  \tilde{\gamma} \left( u\right), \label{eqgamcgam}
\end{equation}
where $c$ is a multiplicative constant independent of direction $u$.
What is less clear is the interpretation of (\ref{eqcgamma}) for directions $%
u$ that do not correspond to equilibrium facets. From the Wulff
construction, one can only conclude that $\gamma \left( u\right) $ is the
minimum fictitious interface free energy that facet $u$ could have without
changing the Wulff shape. However, as shown in Appendix \ref{appwulff},
$\gamma \left( u\right) $ nevertheless has a clear physical interpretation:
It is the surface free energy of a macroscopically flat surface with normal
$u$ but that is, at the microscopic level,\ made of facets from the Wulff
shape. These facets are big enough so that the edge energies are negligible
relative to the surface energies, but small enough so that the interface
still appears macroscopically flat. Thus, the inverse Wulff construction
$\gamma \left( u\right) $ automatically extrapolates the surface free
energies to the \textquotedblleft unstable\textquotedblright\ directions in
a physically plausible fashion.

The determination of the precipitate shape $\mathcal{P}$ from the Monte
Carlo configuration snapshots typically requires the use of a suitable order
parameter to identify which points belong or not to the precipitating phase.
In the special case of precipitates forming from a dilute solid solution on
the parent lattice, one can simply use the species that is dilute (call it
\textquotedblleft D\textquotedblright ) in the solid solution phase as a
marker for the presence of the precipitating phase. One can easily screen out
the solutes in solution by counting, for each D atom, the number of other D
atoms in their vicinity and eliminating any D atom with an insufficient
number of neighbors. The remaining D atom (in concentrated environments)
should nicely trace out the precipitate shape.

When using this approach, it must be verified that the precipitate size used
is sufficiently large to ensure (i) a small Gibbs-Thomson effect \cite{gibbs:thomson,lupis:thermo}
(i.e. changes in chemical potential due to interface curvature) and
(ii) a small ratio of the lattice parameter over the precipiate size.
Note that the method is not sensitive to an overall direction-independent bias
in interfacial free energies that could be due to the Gibbs-Thomson effect,
because the inverse Wulff construction is only used to provide the
relative values of the different interfacial free energies.
Their overall direction-independent scale factor is determined by a separate
calculation on a planar interface (described in the next section) that is
therefore not affected by the Gibbs-Thomson effect.
However, the Gibbs-Thomson effect could still affect the relative stability of two interface orientations
through changes in equilibrium solute concentration, which in turn affect interface structures and their relative excess free energies.
This effect is typically of a smaller magnitude than the isotropic component of Gibbs-Thomson-induced bias,
but must still be carefully investigated.
A second size issue arises from the fact that the lattice parameter places a lower bound on the ``resolution''
of our estimates of precipitate shape. Both the Gibbs-Thomson effect and the ``resolution'' bound
scale as $1/r$, where $r$ is of the order of precipitate ``radius''.

\subsection{Absolute interface free energies}

In order to fix the arbitrary constant $c$ in the inverse Wulff construction
in Equation (\ref{eqgamcgam}) we need to compute the absolute surface free
energy for one facet. We have the freedom to pick a facet that simplifies
the calculations. In this case, taking a facet that does not roughen proves
useful, because the fluctuations in the interface structure are small, so
that little Monte Carlo averaging is needed to obtain converged values.

In general (whether the chosen interface roughens or not), the interface free energy $F_{i}$ can be calculated from the
interface excess energy $E_{i}$ by thermodynamic integration \cite{woodward:nial} of the relation 
$\partial \left( F_{i}/T\right) /\partial \left( 1/T\right) =E_{i}$. To this
effect, one can perform a sequence of lattice gas Monte Carlo simulations,
starting from a perfectly atomically flat interface at $T=0$K, where it is
known that $F=E$, and integrating up to the desired temperature.
Performing this procedure for various supercell periodicities (perpendicular to the
interface), one can infer the contribution of the interface to the free
energy.
Alternatively, one can also use a Gibbs dividing surface construction to subtract
appropriate amount of the free energy of the two bulk phases in contact from
the total free energy of the supercell containing the interface. A natural
choice of dividing surface is one that implies an excess solute
concentration of zero.

In some systems, one can find an equilibrium facet that remains atomically
flat up to the temperature of interest.
This may happen when that interface's low free energy is driven by a low interfacial
energy rather than by a large interfacial entropy, i.e., when
the energy associated with step formation is high.
In this case, the thermodynamic
integration process becomes redundant because there is no configurational
contribution to the interfacial entropy. Hence, one can equivalently
directly calculate the free energy of formation of a sharp interface
(including vibrational contributions) using the effective cluster
interactions at the appropriate temperature and a dividing surface construction.
In this simplified geometry and with a
perfect stoichiometry, the determination of the dividing surface that makes
the excess solute zero is also especially simple. These simplifications turned
out to be possible in the case of the Al$_{3}$Ti precipitates considered
here as an example.

\subsection{Parametrization of the Orientation Dependence}

The interfacial excess free energy provided by Equation (\ref%
{eqcgamma}) is unfortunately not in a very convenient form. First, it is
given numerically in tabular form on a mesh of possible directions. Second,
it is contaminated by noise due to the fact that facets are rarely perfect
(and in fact, are not perfect in general, due to the stabilizing effect of
entropy) and due to ambiguities in defining the interface, as some solute atoms may be misclassified as
part of the precipitate. The resulting interfacial free energies are thus
not guaranteed to satisfy the symmetry constraints imposed by the
crystallography of the problem. These problems can be jointly addressed by
fitting the raw output of Equation (\ref{eqcgamma}) to a small set of
direction-dependent harmonics that are adapted to the known symmetry of the
precipitate's crystal structure. We now describe a simple and general method
to generate appropriate harmonics which exploits a convenient
characterization\ of spherical harmonics as\ polynomials in the components
of a unit vector.

Suitable treatments for special cases already exist in the literature.
Notable examples include the cubic\ harmonics (e.g. \cite%
{altmann:cubharm,muggli:cubharm}) and harmonics for hexagonal symmetry (e.g. 
\cite{altmann:hexharm,hart:cshex}). A very general treatment has already been
presented in the literature \cite{kara:harm}. Although very complete, this treatment does not lend
itself to a simple implementation: The point group and its orientation has
to be identified, not just as a list of symmetry operations, but recognized
by name as one of the known point groups, so that one can lookup the
specific rules applying to that point group. Based on the point group
category found (e.g., cubic or hexagonal), a superset of harmonics is
selected. Then, based on the specific point group found, \textquotedblleft
index rules\textquotedblright\ are applied to eliminate those harmonics that
should vanish by symmetry. This treatment is ideally suited for researchers
wanting to manually construct a basis based on the knowledge of the point
group, as the different cases are nicely classified by point group. However,
a computer program implementing the method would also necessarily contain a
large number of tests and subcases. It would also have to rotate the
symmetries into a standard \textquotedblleft setting\textquotedblright\ to
use the tabulated index rules. Moreover, if one wishes to handle other
point groups that are not special cases of cubic or hexagonal symmetries
(e.g. icosahedral symmetry or other noncrystallographic point groups, which
could occur for quasicrystals), a different superset of harmonics and index
rules must be constructed.

In contrast, we describe here an approach that works directly with the
symmetry operation in matrix form (which are easy to determine) and requires
no classification into categories of point groups. The possibility of having
point groups in different orientations (or \textquotedblleft
settings\textquotedblright ) is automatically handled, with no extra coding
effort. The algorithm only relies on basic linear algebra operations and
handles any point group, not just those for which supersets of harmonics
have already been constructed. The proposed method is related to the one
proposed in \cite{avdw:gce} to generate tensor bases, although additional
steps, provided herein, were needed to formally show that such tensors bases
can be used to generate direction-dependent harmonics and to avoid redundant
harmonics via a projection scheme.

While we outline the method below, a formal algorithm is given in Appendix %
\ref{secalgo}. Polynomials are known to form a complete basis for any
continuous function over a bounded region (e.g. the unit sphere). Hence, in
particular, they form a complete basis for any continuous function defined
over the surface of the unit sphere. Let $u$ be a three-dimensional unit
vector (e.g. $u=\left( \cos \left( \theta \right) \sin \left( \phi \right)
,\sin \left( \theta \right) \sin \left( \phi \right) ,\cos \left( \phi
\right) \right) $).\ Any continuous\ function $f$ of direction $u$ can
therefore be represented as%
\begin{equation}
f\left( u\right) =\sum_{L=0}^{\infty }A^{\left( L\right) }u^{L}
\label{eqexp}
\end{equation}%
where we use the short-hand notation%
\begin{equation}
A^{\left( L\right) }u^{L}\equiv \sum_{i_{1}=1}^{3}\cdots
\sum_{i_{L}=1}^{3}A_{i_{1}\cdots i_{L}}^{\left( L\right)
}\prod_{j=1}^{L}u_{i_{j}} \label{eqshorth}
\end{equation}%
(with $A^{\left( 0\right) }u^{0}$ defined as constant) and where $%
A_{i_{1}\cdots i_{L}}^{\left( L\right) }$ is a rank $L$ tensor that is
symmetric under permutation of the indices (since permutations of the
indices does not change the polynomial we can, without loss of generality,
limit ourselves to such symmetric tensors).\ If the function $f\left(
u\right) $ is constrained by symmetry, such constraints can then be
implemented by restricting the tensors $A^{\left( L\right) }$ to obey
suitable invariance with respect to all symmetry operations in a given point
group \cite{nye:tensor}. As explained in more detail in the Appendix \ref%
{secalgo}, this can be simply accomplished by considering $3^{L}$
noncolinear trial tensors, and obtaining symmetrized tensors by averaging
each trial tensor with all its transformations by each point group symmetry
operation. The desired result is obtained after eliminating colinear
symmetrized tensors.

An additional step is needed because expansion (\ref{eqexp}) is a bit
redundant, since the polynomial $\left( u_{1}^{2}+u_{2}^{2}+u_{3}^{2}\right) 
$ is constant over the unit sphere. This would imply that nonzero
coefficients $A^{\left( L\right) }$ for $L>0$ could give rise to a constant $%
f\left( u\right) $, which is undesirable. This can be avoided by projecting
each $A^{\left( L\right) }$ onto the space orthogonal to tensors giving rise
to polynomials that can be factored as%
\begin{equation}
\left( u_{1}^{2}+u_{2}^{2}+u_{3}^{2}\right) A^{\left( L-2\right) }u^{L-2}
\label{eqfact}
\end{equation}%
for some tensor $A^{\left( L-2\right) }$ of rank $L-2$.\ (It is not
necessary to consider higher powers of $\left(
u_{1}^{2}+u_{2}^{2}+u_{3}^{2}\right) $ in this factorization, because $%
A^{\left( L-2\right) }u^{L-2}$ could include additional $\left(
u_{1}^{2}+u_{2}^{2}+u_{3}^{2}\right) $ factors as a special case.) A simple
algorithm to accomplish the symmetrization and this projection is provided
in Appendix \ref{secalgo}.

The result of this procedure is an expression for the tensor $A^{\left(
L\right) }$ as a sum of $K_{L}$ symmetry-constrained and non-redundant
components $A^{\left( L,k\right) }$: 
\begin{equation*}
A^{\left( L\right) }=\sum_{k=1}^{K_{L}}c_{L,k}A^{\left( L,k\right) }
\end{equation*}%
where the tensors $A^{\left( L,k\right) }$ are fixed and determined by
symmetry while the coefficients $c_{L,k}$ are completely unrestricted. Upon
substitution into (\ref{eqexp}) we obtain a symmetry-constrained expansion:%
\begin{equation}
f\left( u\right) =\sum_{L=0}^{\infty }\sum_{k=1}^{K_{L}}c_{L,k}A^{\left(
L,k\right) }u^{L}.  \label{eqexp2}
\end{equation}

It is instructive\ to verify that expansion (\ref{eqexp2}) coincides (apart
from an inconsequential linear transformation) with spherical harmonics when
no symmetry constraints are imposed. The easiest way to see this is to
compare (\ref{eqexp2}) with the eigenfunction of the Schr\"{o}dinger
equation for some\ spherically symmetric potential selected so that the
eigenfunctions involve polynomials. We can use any convenient radial
potential because we only focus on the angular part.\ Consider a spherically
symmetric harmonic potential, whose eigenstates are polynomials times a
spherically symmetric Gaussian. The Gaussian is constant over the unit
sphere, so we are left with only a polynomial as the angular dependence.\
Moreover, it is well-known that the order of that polynomial is equal\ to $%
n_{1}+n_{2}+n_{3}$, the sum of three principal quantum numbers of the
harmonic oscillator along each dimension. This sum is also (up to a constant
scaling and shift)\ the energy of the system. It follows that there is a
direct correspondence between all terms in (\ref{eqexp2}) sharing the same
value of $L$ and all eigenfunctions sharing the same energy. The different
terms sharing the same $L$ thus correspond to eigenstates with different
angular momentum projections.\ We can verify that the number of terms (in
the case of a spherically symmetric potential) matches the number of
spherical harmonics for a given value of $L$. Indeed, the number of distinct
terms of total power $L$ in a polynomial in $v$ variables is $\binom{L+2}{v-1%
}$. In the present case $\binom{L+2}{3-1}=\left( L+2\right) \left(
L+1\right) /2$.\ From that number, we subtract the dimension of the subspace
of polynomials that factor as $\left( u_{1}^{2}+u_{2}^{2}+u_{3}^{2}\right) $
times a polynomial of order $L-2$, we obtain $\binom{L+2}{2}-\binom{L}{2}%
=\left( L+2\right) \left( L+1\right) /2-L\left( L-1\right) /2=2L+1$, exactly
the number of spherical harmonics associated with angular momentum is $L$.
Hence the dimension of the space spanned by the spherical harmonics for a
given $L$ is the same as the dimension of the space spanned by our
polynomials. Both spaces include polynomials of order $L$ on the unit sphere
that are not colinear and it follows that both bases must span the same
space. Hence both expansions, truncated to the same $L$, span the same space.

The algorithm proposed above has been implemented within the\ Alloy
Theoretic Automated Toolkit (ATAT) \cite%
{avdw:maps,avdw:atat,avdw:atat2,avdw:atatcode} as the \textquotedblleft
gencs\textquotedblright\ code, documented in Section \ref{secsupp}.
For convenience, the coefficients $A^{\left( L\right) }$ (up to $L=6$) 
for all crystallographic and selected noncrystallographic point groups can also be found in
electronic form in the Supplementary Material ({\tt suppmat.txt}). Figure \ref{ptgrpfig} represents these harmonics
graphically for each of the crystallographic point groups.\ Figure \ref%
{nonxtalfig} shows the result of a similar exercise for selected
noncrystallographic point groups, which could be useful, for instance, to handle the case
of quasicrystals precipitating out of a liquid.

\begin{figure}
\centerline{\includegraphics{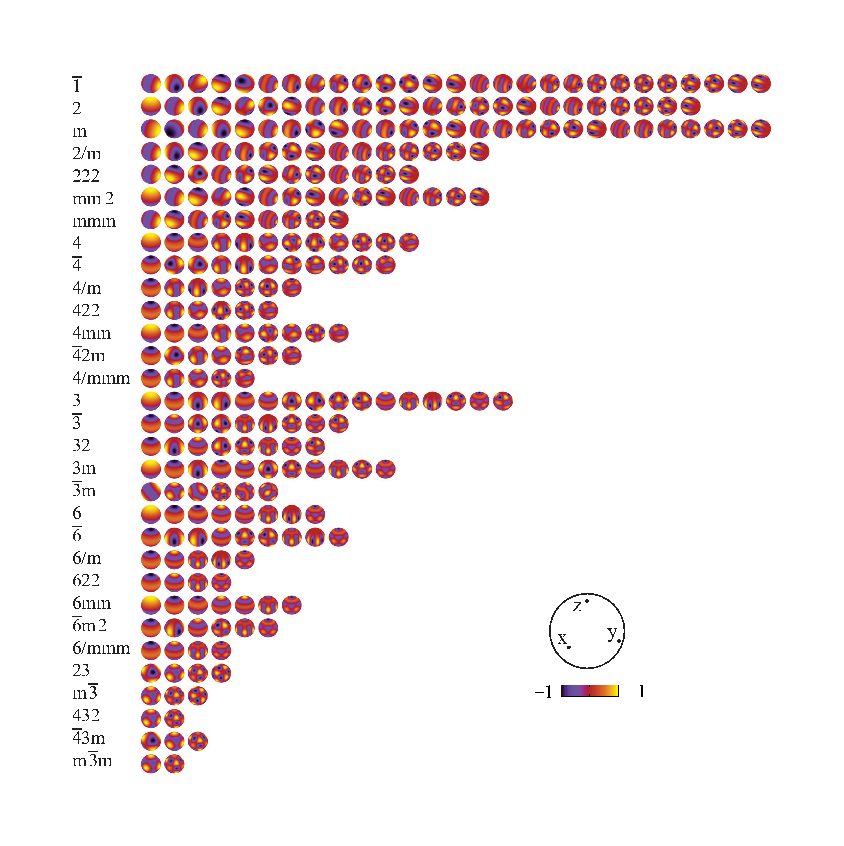}}
\caption{\label{ptgrpfig} (Color online) Symmetry-adapted spherical harmonics
for each crystallographic point groups (up to L=6).
Group 1 (with 48 harmonics) is omitted.
The inset shows the location of the Cartesian axes  on the unit sphere
(where directions of highest symmetry are aligned, whenever possible)
and the color scheme used to represent the function.
Unique axis (when appropriate) is chosen to be $z$.}
\end{figure}%

\begin{figure}
\centerline{\includegraphics{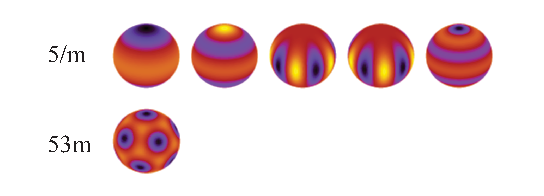}}
\caption{\label{nonxtalfig} (Color online) Symmetry-adapted spherical harmonics
for selected noncrystallographic point groups (up to L=6),
represented using the same conventions as in Figure \ref{ptgrpfig}.
}
\end{figure}%

\section{Results}

\subsection{Cluster expansion}

\begin{figure}
\centerline{\includegraphics{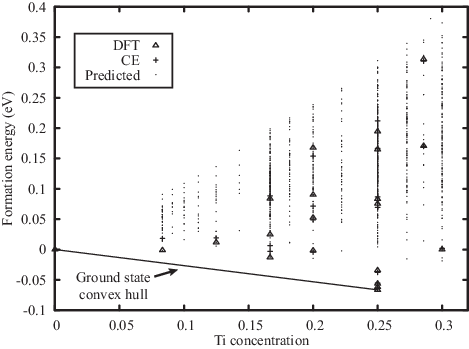}}
\caption{\label{figgs}Calculated DFT formation energies (triangles) and corresponding energies from the fitted cluster expansion (crosses).
Dots mark predicted energies of structures not included in the fit. The line marks the convex hull of the ground state energies.}
\end{figure}

\begin{figure}
\centerline{\includegraphics{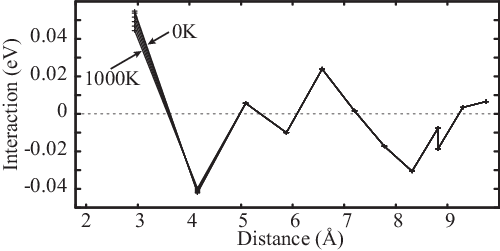}}
\caption{\label{figteci}Effective interaction $J_{\cdots}$ as a function of distance.}
\end{figure}

Since our focus is in the equilibrium between an Al-rich solid solution and
Al$_{3}$Ti precipitate, the range of composition sampled during the cluster
expansion construction process was restricted to less than 30 atomic percent
Ti. Agreement between the ground state convex hulls from the raw DFT
energies and the energies obtained from the cluster expansion (see Figure %
\ref{figgs})\ was enforced in the range of 0 to 25 atomic percent Ti.\
Restricting the composition range in this fashion drastically improves the
convergence of the cluster expansion. The cluster expansion construction
process necessitated the calculation of the formation of energy, using ab
initio methods, of $21$ structures ranging from $1$ to $12$ atoms per unit
cell. The ground state search was extended up to $12$ atoms per unit
cells. The resulting cluster expansion exhibits\ a mean square error of $15$
meV.\ In this system, only pair interactions were found to be necessary, as
determined from a cross-validation analysis \cite{avdw:maps}. These
interactions are depicted in Figure \ref{figteci}.

\begin{figure}
\centerline{\includegraphics{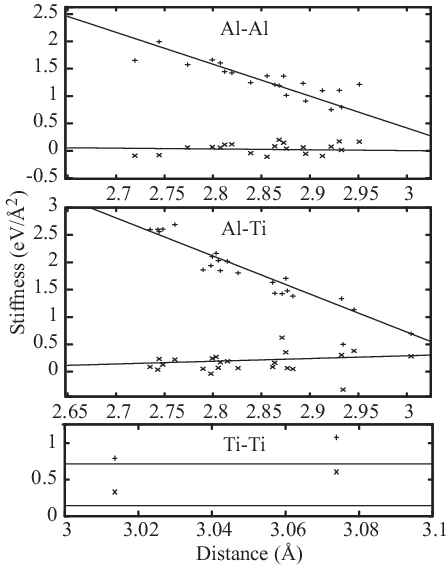}}
\caption{\label{figsvsl}Stiffness vs. Length relationship obtained from a fit to three structures
and used to efficiently calculate the phonon spectrum for a large number of structures.
Stiffness data for bond stretching and bending are marked by $``+''$ and $``\times''$, respectively.
The resulting linear fits are used in subsequent phonon calculations.}
\end{figure}

The temperature dependence of these interactions was calculated via the
transferable force constant approach (also called the \textquotedblleft bond
stiffness vs bond length\textquotedblright\ approach)
\cite{avdw:vibrev,avdw:pd3v}. Three structures were used in the fit of the force
constants (pure Al, Al$_{3}$Ti in the DO$_{23}$ structure and a metastable
Al$_{7}$Ti$_{3}$ structure with 10 atoms per unit cell, chosen for its small
size and the presence of Ti-Ti bounds) and each were considered at their
equilibrium lattice parameters at 0K as well as under a linear strain of $2\%
$. The resulting length-dependence of the force constants is illustrated in
Figure \ref{figsvsl}, along with the input ab initio stiffness data. These
ab initio phonon calculations were performed using supercells ranging from
32 to 48 atoms, which is sufficient given the nearest-neighbor nature of the
transferable force constants. These transferable length-dependent force
constants were then used to calculate phonon spectra for all 21 structures
used in the cluster expansion construction. A cross-validation analysis indicated that
the configuration-dependence of the phonon free energy can be captured using
only the $3$ nearest neighbors pairs in the cluster expansion. Although the
temperature-dependence of the interactions (Figure \ref{figteci}) may appear
small, it nevertheless has a significant impact on the temperature scale of
thermodynamics of the system. In precipitation calculations neglecting the
effect of phonons (not described here, for conciseness), the precipitate
exhibited flat facets up to around $1500$K, while roughening occurs below $900
$K when vibrational effects are included.

\subsection{Monte Carlo simulations}

Canonical Monte Carlo simulations were performed with the \textquotedblleft
emc2\textquotedblright\ code \cite{avdw:emc2} included in ATAT which
implements a standard Metropolis algorithm. Two types of simulations were
used: (i) planar interface simulations, to obtain one absolute excess free
energy and (ii) and precipitate shape equilibration simulations, to obtain
the relative excess free energies for all interface directions.

The planar interface simulations were performed in a $12\times 12\times n$
supercells of the cubic conventional cell of fcc, with $n=18,24,30$.
The supercell consisted of 1/2 Al$_{3}$Ti and 1/2 pure Al, resulting in two $\{001\}$ interfaces. During the
thermodynamic integration runs, we observed that the $\left\{ 001\right\} $
interfaces remained atomically flat up to $900$K, thus suggesting that
configurational contribution to the interfacial excess are negligible and
that thermodynamic integration is unnecessary. It was verified that this
finding was not merely an artifact of insufficient equilibration, by
deliberately starting the simulation with an excess solute dissolved in the
host phase and observing that these solutes rapidly attach to the surface,
one layer at a time and remain in place for the duration of the
simulation.\ Given this behavior, we report here the excess free energy of
a\ perfectly flat $\left\{ 001\right\} $ interface obtained directly from
the cluster expansion (which agrees, within numerical integration noise,
with the full thermodynamic integration results). The interfacial excess
free energy of the $\left\{ 001\right\} $ interfaces at $900$K is $240$\ mJ/m%
$^{2}$, using a dividing surface that makes the interfacial excess solute (Ti) vanish.
Areas are calculated assuming the lattice parameter of pure Al,
corresponding to the limit of small\ coherent precipitates in a dilute solid
solution. This is the relevant limit since the concentration of Ti solutes
in the host phase (fcc Al) was found to be less than $10^{-4}$.

To verify convergence of the result with respect to precipitate size, we performed
simulations for a range of precipitate sizes.
The simulation cells considered were $n\times n\times n$ supercells of the cubic conventional cell of fcc
with $n=24, 36, 48, 60$. In each case, the precipitate was a parallelepided occupying $2/3$ of
the simulation supercell along each direction (to ensure that the precipitate does not
interact with its periodic images). These correspond to simulations
involving from $\sim 55,000$ to $864000$ atoms with precipitates containing from $\sim 12,000$ to $\sim 225,000$ atoms.
In each case, the precipitate was first equilibrated for at least $10,000$ Monte Carlo passes at
the higher temperature of $1800$K (to speed up to process) before being
equilibrated at the final temperature of interest $900$K for at least $16,000$
passes. The inverse Wulff construction were performed on a $10$ snapshots separated by
$1000$ Monte Carlo passes and averaged to yield the data reported here.
In a medium-sized supercell ($36 \times 36 \times 36$), it was verified that the simulation equilibrated to similar shapes even when starting
from different initial solute shape: (i) an octahedron made of $\{111\}$ facets and with longest axis of length $32$ and
(ii) a $24\times 24\times 12$ parallelepiped. The resulting equilibrium shape
did not detectably change upon further equilibration for another $50,000$ Monte Carlo passes at $900$K.

\begin{figure}
\centerline{\includegraphics{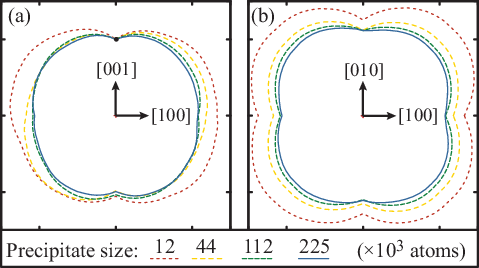}}
\caption{\label{convsizefig}
(Color online) Convergence of the direction-dependence of the interfacial free energy $\tilde{\gamma}(u)$ as
a function of precipitate size along two planar cross-sections. $\tilde{\gamma}(u)$ is normalized so that $\tilde{\gamma}((0,0,1))=1$, as indicated by the black circle.
These plots show the unprocessed data, prior to a fit to harmonics (hence deviations from symmetry can be seen).
}
\end{figure}

Figure \ref{convsizefig} show the convergence of the direction-dependence of the interfacial free energies
obtained directly from the precipitate shape prior to a fit with harmonics.
At small precipitate sizes, one can both see a systematic bias (due to the Gibbs-Thomson effect) and
significant random noise (due to the resolution limit implied by the lattice parameter).
The difference between the results obtained with the two largest precipitate sizes never differ by more than 5\%,
which can be taken as an upper bound on the magnitude of the errors.

\subsection{Interfacial excess free energies}

\begin{figure}
\centerline{\includegraphics{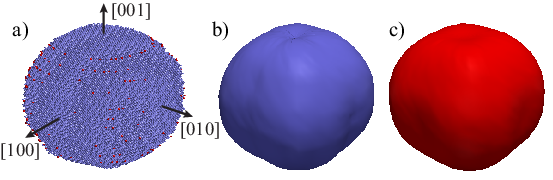}}
\caption{\label{figprectoharm}
(Color online) (a) Equilibrated Al$_{3}$Ti DO$_{23}$ precipitate (at $900$K) consisting of
approximately $225,000$ atoms (only Ti atoms are shown). Atoms identified as solute are marked in red. Note the combination
of atomically flat $\left\{ 001\right\} $ facets and roughened interfaces in
other directions.
(b) Inverse Wulff construction from the precipitate shape that provides the
relative interfacial excess free energy as a function of direction (units
are dimensionless).
(c) Symmetry-constrained harmonic expansion fitted to the data in b).
}
\end{figure}

\begin{figure}
\centerline{\includegraphics{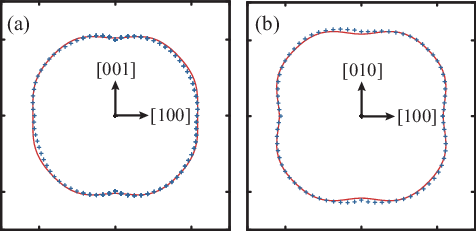}}
\caption{\label{fitshapefig}
(Color online) Panels (a) and (b) overlap the raw data (blue crosses) from the inverse Wulff construction
(from Figure \ref{convsizefig} for the largest precipitate size) and the least squares fit of
a harmonic expansion (red continuous curve), along two planar cross-sections.
}
\end{figure}

The resulting equilibrium precipitate shape was fed to the inverse Wulff
construction (Equation (\ref{eqcgamma})), after the few solute atoms present
in the Al matrix were eliminated by removing all Ti atoms no more than $2$ Ti neighbors within a $5.1$ \AA\ radius.
As illustrated in Figure \ref{figprectoharm}, the Wulff
plot obtained in this fashion is slightly noisy, but a least square fit to
symmetry-constrained harmonics yield a well-behaved Wulff plot
$\tilde{\gamma} \left( u\right) $ obeying the underlying symmetry of the DO$_{23}$ phase ($4/mmm$).
Cross-sections of the inverse Wulff construction are shown in Figure \ref{fitshapefig} along with their
corresponding harmonic fits.

When facetting occurs, the Wulff plot $\gamma \left( u\right) $ contains
nonsmooth cusps in the direction of the facets which may be difficult to
represent with only a few smooth harmonics. However, it is easy to generate
as many data points (i.e. directions $u$) as needed and as many harmonics as
needed to alleviate this potential problem. In the present setting,
including harmonics up to $L=8$ and sampling the unit sphere on a grid of $40$
different latitudes and $80$ different longitudes was found to be
sufficient for this purpose. It was found helpful to increase the weight of points near the cusp
during the fit to ensure it is better reproduced. In the present system, a direction at an angle $\phi$
from the $[001]$ axis was given a weight of $\omega + (1-\omega)\sin \phi$ with $\omega=0.2$.
(A uniform weight would have been of the form $\sin \phi$, but the additional $\omega$ term increases
the weight near the cusp at $\phi=0$.)

The resulting calculated interfacial excess free energies are
reported in Table \ref{tabharm}, as a linear combination of these symmetry-constrained
harmonics. The statistical noise (arising from the random fluctuations visible
in Figure \ref{fitshapefig}) introduce errors in the calculated free energies that 
are less than 1\% and are thus negligible relative the precipitate size convergence errors (at most 5\%).
Excess free energies along selected directions, calculated from these harmonics,
are also reported in Table \ref{tabdir}.

\begin{table}
\caption{\label{tabharm}Harmonic expansion (Equations (\ref{eqexp2}) and (\ref{eqshorth}))
of the calculated interfacial excess free energies.
Harmonics are expressed in terms of the components $\left( x,y,z\right) $ of a unit vector $u$.
The number of significant digits reported reflects the standard errors of the statistical regression (1--5 mJ/m$^2$).
Finite size effects introduce an additional error of at most 5\%.
}
\begin{tabular}{ccl}
\hline
Coefficient & Value & Harmonics\\
$c_{L,k}$& (mJ/m$^{2}$) & $A^{(L,k)}u^{L}$\\
 \hline\hline
$c_{0,1}$ & 282 & $1$ \\ \hline
$c_{2,1}$ & 18  & $0.408x^{2}+0.408y^{2}-0.816z^{2}$ \\ \hline
$c_{4,1}$ & -59 & $0.396x^{4}-1.730x^{2}y^{2}+0.396y^{4}-0.649x^{2}z^{2}-0.649y^{2}z^{2}+0.216z^{4}$ \\ \hline
$c_{4,2}$ & 2   & $1.279x^{2}y^{2}-1.279x^{2}z^{2}-1.279y^{2}z^{2}+0.426z^{4}$\\ \hline
$c_{6,1}$ & 14  & $0.154x^{6}-0.421x^{4}y^{2}-0.421x^{2}y^{4}+0.154y^{6}-1.894x^{4}z^{2}+$ \\ 
          &     & $+5.050x^{2}y^{2}z^{2}-1.894y^{4}z^{2}+1.052x^{2}z^{4}+1.052y^{2}z^{4}-0.140z^{6}$ \\ \hline
$c_{6,2}$ & 6   & $0.557x^{4}y^{2}+0.557x^{2}y^{4}-0.557x^{4}z^{2}-6.680x^{2}y^{2}z^{2}-0.557y^{4}z^{2}+$ \\ 
          &     & $+1.670x^{2}z^{4}+1.670y^{2}z^{4}-0.223z^{6}$ \\ \hline
$c_{8,1}$ & -73 & $0.113x^8-2.011x^6y^2+4.640x^4y^4-2.011x^2y^6+0.113y^8-1.141x^6z^2+$ \\
          &     & $+2.320x^4y^2z^2+2.320x^2y^4z^2-1.141y^6z^2+2.465x^4z^4-4.640x^2y^2z^4+$ \\
          &     & $+2.465y^4z^4-0.677x^2z^6-0.677y^2z^6+0.0483z^8$ \\ \hline
$c_{8,2}$ & -9  & $1.470x^6y^2-3.956x^4y^4+1.470x^2y^6-1.470x^6z^2+1.686x^4y^2z^2+$ \\
          &     & $+1.686x^2y^4z^2-1.470y^6z^2+3.394x^4z^4-3.371x^2y^2z^4+3.394y^4z^4+$ \\
          &     & $-1.133x^2z^6-1.133y^2z^6+0.0809z^8$ \\ \hline
$c_{8,3}$ & -27 & $1.224x^4y^4-7.345x^4y^2z^2-7.345x^2y^4z^2+1.224x^4z^4+14.689x^2y^2z^4+$ \\
          &     & $+1.224y^4z^4-1.469x^2z^6-1.469y^2z^6+0.105z^8$ \\ \hline
\end{tabular}
\end{table}

\begin{table}
\caption{\label{tabdir}Calculated interfacial excess free energies along selected directions
(within an accuracy of 5\%, due to finite size effects).
}
\begin{tabular}{ccc}
\hline
Direction & & Interfacial Free energy (mJ/m$^2$) \\
\hline
\{0 0 1\} & & 245 \\
\{1 0 0\} & & 259 \\
\{1 1 0\} & & 268 \\
\{1 0 1\} & & 206 \\
\{1 1 1\} & & 227 \\
\hline
\end{tabular}
\end{table}

\section{Conclusion}

Apart from a scaling constant,
the complete orientation-dependence of interface free energies can be inferred from
equilibrium precipitate shapes via the inverse Wulff construction.
The scaling constant can be recovered
from a calculation of the excess free energy of a single, conveniently chosen, planar interface,
from a thermodynamic integration procedure starting from a sharp interface at absolute zero.

The present work goes beyond these simple realizations along many key aspects.
We employ the cluster expansion formalism in its most general coarse-graining form to efficiently model the system's
energetics in a way that includes the effect of lattice vibrations without necessitating explicit modeling of the
atomic dynamics throughout the simulation.
This approach provides sufficient efficiency to reach the simulation system sizes and the equilibration times needed to
obtain properly equilibrated precipitates of a size sufficient to enable the reliable determination of their shapes.
The method is able to account for essentially all sources of entropy (arising
from phonons, bulk point defects, as well as interface roughness) and is
thus able to transparently handle both atomically smooth and rough
interfaces. This feature is illustrated by an application to precipitates in the Al-Ti alloy system.

We also address the conceptual issue that some interface directions do not
appear on the Wulff shape when facetting occurs. We show how the
interfacial free energy surface can be naturally extended into the
``masked'' regions of the Wulff plot in a
way that (i) preserves the predicted equilibrium shape and (ii) has a
natural geometric interpretation.

We provide symmetry-adapted harmonic bases (both in the form of a simple algorithm and as explicit expressions
for all crystallographic point groups) to represent the resulting orientation-dependent interfacial free energies.
The same bases could prove more generally useful in a variety of other applications,
for instance to represent the anisotropy of the so-called constituent strain
elastic energy.

\section*{Acknowledgements}

This work is supported by XSEDE computing resources and by the National
Science Foundation under Grant No. DMR-0907669.

\appendix

\section{Harmonic generation method}

\subsection{Definitions}

\label{secalgo}Let us first define a few convenient symbols.

\begin{itemize}
\item Let $S$ denote a $3\times 3$ matrix representing a point symmetry
operation in Cartesian coordinates and let the corresponding\ function $%
S\left( A^{\left( L\right) }\right) $ applied to a tensor $A^{\left(
L\right) }$ of rank $L$ be defined as:%
\begin{equation*}
\left[ S\left( A^{\left( L\right) }\right) \right] _{j_{1},\ldots
,j_{L}}=\sum_{i_{1}=1}^{3}\cdots \sum_{i_{L}=1}^{3}A_{i_{1},\ldots
,i_{L}}^{\left( L\right) }\prod_{k=1}^{L}S_{i_{k}j_{k}}
\end{equation*}%
and let $\mathcal{S}$ denote a set of such matrices that defines the point
group of interest.

\item Let $P$ denote a permutation vector (i.e. an $L$-dimensional vector
containing all the number $\left\{ 1,\ldots ,L\right\} $ not necessarily in
increasing order) and let the function $P\left( \cdot \right) $ be defined
as: 
\begin{equation*}
\left[ P\left( A^{\left( L\right) }\right) \right] _{j_{1},\ldots
,j_{L}}=A_{i_{P\left( 1\right) },\ldots ,i_{P\left( L\right) }}^{\left(
L\right) }
\end{equation*}%
and let $\mathcal{P}^{\left( L\right) }$ denote the set of all such
permutations for a given $L$.

\item Let $\#$ denote the number of elements in a set.

\item Let $C^{\left( L,1\right) },\ldots ,C^{\left( L,M\right) }$ be a set
of rank $L$ tensors\ defining linear constraints on the generated tensor
basis, i.e. the $A^{\left( L,k\right) }$ generated must be orthogonal to all 
$C^{\left( L,m\right) }$, according to the inner product 
\begin{equation}
A^{\left( L,k\right) }\cdot C^{\left( L,m\right) }\equiv
\sum_{i_{1}=1}^{3}\cdots \sum_{i_{L}=1}^{3}A_{i_{1},\ldots ,i_{L}}^{\left(
L,k\right) }C_{i_{1},\ldots ,i_{L}}^{\left( L,m\right) }.  \label{eqdotprod}
\end{equation}%
(If $M=0$, no constraints are imposed.)

\item Let $\mbox{vec}\left( A^{\left( L\right) }\right) $ denote a
vectorization of the tensor $A^{\left( L\right) }$ (i.e. a $3^{L}$%
-dimensional column vector containing all elements of the tensor $A^{\left(
L\right) }$) and let $\mbox{unvec}\left( \cdot \right) $ denote the reverse
operation.
\end{itemize}

\subsection{Algorithm}

Our algorithm for generating a basis for tensors of rank $L$ obeying
symmetric constraints (defined by a point group $\mathcal{S}$), indices
permutation invariance constraints (defined by the set $\mathcal{P}^{\left(
L\right) }$) and some linear constraints (defined by a basis of symmetric
(under index permutations)\ tensors $C^{\left( L,1\right) },\ldots
,C^{\left( L,M\right) }$) is then as follows:

\begin{enumerate}
\item If $M>0$, define $B$ to be the nonzero and non linearly dependent
columns of the matrix:%
\begin{equation*}
B=\left[ \mbox{vec}\left( C^{\left( L,1\right) }\right) ,\ldots ,\mbox{vec}%
\left( C^{\left( L,M\right) }\right) \right] .
\end{equation*}

\item Set $k=0$

\item Consider a set of distinct trial tensors $\tilde{A}^{\left( L,t\right)
}$ for $t=1,\ldots ,3^{L}$, each consisting of a single element set to $1$,
with all remaining elements set to $0$ and set 
\begin{equation*}
A^{\left( L,t\right) }=\frac{1}{\#\mathcal{P}}\sum_{P\in \mathcal{P}}P\left( 
\tilde{A}^{\left( L,t\right) }\right) .
\end{equation*}%
(For added efficiency, one can limit the trial tensors to those whose
nonzero element $\tilde{A}_{i_{1},\ldots ,i_{L}}$ obeys $i_{1}\leq i_{2}\leq
\ldots \leq i_{L}$. Also, it is clear that the sum over permutations $%
\mathcal{P}$ is just equivalent to setting to $1$ all elements of the tensor 
$A^{\left( L,t\right) }$ equivalent to the nonzero element of $\tilde{A}%
^{\left( L,t\right) }$ under index permutations.)

\begin{enumerate}
\item For each trial tensor $A^{\left( L,t\right) }$, set%
\begin{equation*}
Q=\mbox{unvec}\left( \left( I-B\left( B^{T}B\right) ^{-1}B^{T}\right) %
\mbox{vec}\left( A^{\left( L,t\right) }\right) \right)
\end{equation*}%
(or simply $Q=A^{\left( L,t\right) }$ if $M=0$).

\item Calculate%
\begin{equation*}
\bar{Q}=\frac{1}{\#\mathcal{S}}\sum_{S\in \mathcal{S}}S\left( Q\right)
\end{equation*}

\item If $\bar{Q}$ is nonzero (within machine numerical precision) and (for $%
k>0$) not linearly dependent with the $\left[ \bar{A}^{\left( L,1\right)
},\ldots \bar{A}^{\left( L,k\right) }\right] $, then increment $k$ and set $%
\bar{A}^{\left( L,k\right) }=\bar{Q}$.
\end{enumerate}

\item Finally, set $K_{L}=k$ and orthogonalize (and normalize) the element
of $\left[ \bar{A}^{\left( L,1\right) },\ldots \bar{A}^{\left(
L,K_{L}\right) }\right] $ using the Gram-Schmidt procedure.
\end{enumerate}

The harmonics of order up to $L^{\max }$ are then generated by calling the
above routine for $L=1,\ldots ,L^{\max }$, setting $M=0$ if $L\leq 2$ and
otherwise setting the constraints $C^{\left( L,k\right) }$ to be%
\begin{eqnarray*}
C^{\left( L,k\right) } &=&\sum_{P\in \mathcal{P}}P\left( \tilde{C}\right) \\
\text{with }\tilde{C}_{i_{1},\ldots ,i_{L}} &=&\delta _{i_{1,}i_{2}}\tilde{A}%
_{i_{3},\ldots ,i_{L}}^{\left( L-2,k\right) }
\end{eqnarray*}%
where the $\tilde{A}^{\left( L-2,k\right) }$ are also generated with the
above routine,\ called with rank $L-2$, the same point group $\mathcal{S}$,
the permutation set $\mathcal{P}^{\left( L-2\right) }$ and $M=0$ (no
constraints $C^{(L-2,k)}$).

Although it is not explicit in the notation above, it is clear that
efficiency improvements (in storage and computational requirements) are
possible by exploiting, at each step,\ the symmetry of all tensors
considered under permutation of their indices. However, for the basis sizes
we considered in this paper, we found such
optimization to be unnecessary. It is also interesting to note that, since
all operations (except for the last normalization step)\ yield tensors with
rational elements, one can\ use an exact rational representation for the
coefficients to obtain an analytic (rather than numerical) expression for
the harmonics. We did not find this to be necessary, however, and, in fact,
harmonics are often reported in numerical form \cite%
{altmann:cubharm,muggli:cubharm}.

\subsection{Proof of the validity of the symmetrization technique}

It is instructive to see why the method used for symmetrization in the
algorithm of Section \ref{secalgo} actually works. For an arbitrary trial
tensor $A^{\left( L,t\right) }$ we can verify that the symmetrized tensor%
\begin{equation*}
\tilde{Q}=\frac{1}{\#\mathcal{S}}\sum_{S\in \mathcal{S}}S\left( A^{\left(
L,t\right) }\right)
\end{equation*}%
obeys $T\left( \tilde{Q}\right) =\tilde{Q}$ for any operation $T\in \mathcal{%
S}$. Indeed, calculate%
\begin{equation*}
T\left( \tilde{Q}\right) =\frac{1}{\#\mathcal{S}}\sum_{S\in \mathcal{S}%
}T\left( S\left( A^{\left( L,t\right) }\right) \right) =\frac{1}{\#\mathcal{S%
}}\sum_{\tilde{S}\in \mathcal{S}}\tilde{S}\left( A^{\left( L,t\right)
}\right) =\tilde{Q}
\end{equation*}%
where the second equality holds because $T\left( S\left( \cdot \right)
\right) $ is just another operation in $\tilde{S}\left( \cdot \right) \in 
\mathcal{S}$ and two distinct $S\left( \cdot \right) $ cannot be mapped onto
the same symmetry operation by applying $T\left( \cdot \right) $, since each
element of a group admits an inverse. Since the symmetrization procedure is
a linear projection, choosing the trial tensors $A^{\left( L,t\right) }$ so
that they form an orthogonal basis is sufficient to generate a basis for the
space of symmetrized tensors.

A similar argument holds for invariance under permutations $P$ of the
indices. Finally, note that applying a point group operation $S\left( \cdot
\right) $ to a tensor $A^{\left( L\right) }$ that is invariant to indices
permutations yields a tensor with the same property:%
\begin{eqnarray*}
\left[ P\left( S\left( A^{\left( L\right) }\right) \right) \right]
_{j_{1},\ldots ,j_{L}} &=&\left[ S\left( A^{\left( L\right) }\right) \right]
_{j\left( P_{1}\right) ,\ldots ,j\left( P_{L}\right)
}=\sum_{i_{1}=1}^{3}\cdots \sum_{i_{L}=1}^{3}A_{i_{1},\ldots ,i_{L}}^{\left(
L\right) }\prod_{k=1}^{L}S_{i_{k}j_{P\left( k\right) }} \\
&=&\sum_{i_{1}=1}^{3}\cdots \sum_{i_{L}=1}^{3}A_{i_{P\left( 1\right)
},\ldots ,i_{P\left( L\right) }}^{\left( L\right)
}\prod_{k=1}^{L}S_{i_{P\left( k\right) ,}j_{P\left( k\right) }} \\
&=&\sum_{i_{1}=1}^{3}\cdots \sum_{i_{L}=1}^{3}A_{i_{P\left( 1\right)
},\ldots ,i_{P\left( L\right) }}^{\left( L\right)
}\prod_{k=1}^{L}S_{i_{k,}j_{k}} \\
&=&\sum_{i_{1}=1}^{3}\cdots \sum_{i_{L}=1}^{3}A_{i_{1},\ldots
,i_{L}}^{\left( L\right) }\prod_{k=1}^{L}S_{i_{k,}j_{k}}=\left[ S\left(
A^{\left( L\right) }\right) \right] _{i_{1},\ldots ,i_{L}}
\end{eqnarray*}%
where we have used the fact re-ordering the sums or the product has no
effect and the invariance of $A^{\left( L\right) }$ under permutation. This
shows that symmetrizing the tensor after making it invariant to indices
permutations does not undo the permutation invariance.

\section{Interpretation of nonequilibrium excess free energies}

\label{appwulff}Let $u$ be a unit vector and let $\gamma \left( u\right) $
denote the surface (free) energy for an interface with normal $u$. Consider
an interface that appears macroscopically flat with normal $u_{0}$ and unit
area $A_{0}=1$ but that is, microscopically made of $K$ different facets of
orientations $u_{1},\ldots ,u_{K}$ with corresponding areas $A_{1},\ldots
,A_{k}$. We assume that these facets are big enough that the edge energies
are negligible relative to the surface energies, but small enough that the
interface still appears macroscopically flat. Our goal is to express the $%
\gamma \left( u_{0}\right) $ in terms of $u_{0},u_{1},\ldots ,u_{K}$\ and $%
\gamma \left( u_{1}\right) ,\ldots ,\gamma \left( u_{K}\right) $.

The first step is to solve for the $A_{1},\ldots ,A_{K}$. To this effect,
consider a uniform fictitious \textquotedblleft field\textquotedblright\ $f$
traversing the surface $u_{0}$ and observe that this\ flux must be equal to
the flux traversing the facetted surface made of orientations $u_{1},\ldots
,u_{K}$:

\begin{equation*}
A_{0}u_{0}\cdot f=\sum_{k=1}^{K}A_{k}u_{k}\cdot f
\end{equation*}%
Since this must hold for any constant flux $f$ and since $A_{0}=1$ by
convention, we have the vector identity $u_{0}=\sum_{k=1}^{K}A_{k}u_{k}$
which can be written in matrix form as $Ua=u_{0}$ where $U=\left[
u_{1},\ldots ,u_{k}\right] $ and $a=\left[ A_{1},\ldots ,A_{k}\right] ^{T}$.
If $K=3$, $U$ is a $3\times 3$ matrix that is necessarily invertible (for
otherwise some facets would be redundant). If $K=2$, the problem can be
reduced to a two-dimensional problem by a change of coordinates and $U$ is $%
2\times 2$ matrix that is invertible. The $K=1$ case is trivial. We can then
generally solve for the $A_{k}$ via $a=U^{-1}u_{0}$. The effective surface
energy is then%
\begin{equation}
\gamma \left( u_{0}\right) =\sum_{k=1}^{K}\gamma \left( u_{k}\right)
A_{k}=g^{T}a=g^{T}U^{-1}u_{0}  \label{eqgu0}
\end{equation}%
where $g=\left[ \gamma \left( u_{1}\right) ,\ldots ,\gamma \left(
u_{K}\right) \right] ^{T}$.

We can obtain the same answer via a simple geometric construction. Let $x$
be the point of intersection of the facets of the Wulff shape associated
with $u_{1},\ldots ,u_{K}$. This point $x$ can be found by solving the
system of equations $u_{k}\cdot x=\gamma \left( u_{k}\right) $ for $%
k=1,\ldots ,K$. which can be written in matrix form as $U^{T}x=g$, using the
earlier notation. Hence, $x=\left( U^{T}\right) ^{-1}g$. Now consider a plane
with normal $u_{0}$ intersecting $x$. Its distance from the origin is given
by 
\begin{equation*}
x^{T}u_{0}=\left( \left( U^{T}\right) ^{-1}g\right)
^{T}u_{0}=g^{T}U^{-1}u_{0}
\end{equation*}%
which is exactly the same as $\gamma \left( u_{0}\right) $ given by (\ref%
{eqgu0}). This implies that the inverse Wulff construction automatically
extrapolates the surface free energies to the \textquotedblleft
unstable\textquotedblright\ directions to reproduce the energy of a
microscopically facetted surface made of equilibrium facets that are large
enough to make edge energies negligible.

\section{Supplementary Material: Using the gencs code}
\label{secsupp}

The code takes, as an input, either a point group (specified via generators)
or a structural information from which it determines the point symmetry
automatically. It outputs the harmonics in a file, in form that is easy to
read into a computer code and outputs the harmonics in human-readable form
on the standard output. Some of the file formats contain extraneous items
not needed for harmonic generation \emph{per se} (marked in italics below),
but that are included to ensure compatibility with other portions of the
ATAT package.

\subsection{Input files}

By default, the code reads in structural information (from the \texttt{lat.in%
} file by default --- a alternate file name can be specified with the 
\texttt{-l} option) and determines the point group automatically. The lat.in
file has the following format.

\begin{enumerate}
\item First, the coordinate system $\vec{a}$,$\vec{b}$,$\vec{c}$ is
specified, either as%
\[
\lbrack a]~[b]~[c]~[\alpha ]~[\beta ]~[\gamma ] 
\]%
\textbf{or} in terms of Cartesian coordinates, one axis per line:%
\[
\begin{array}{ccc}
\left[ a_{x}\right] & \left[ a_{y}\right] & \left[ a_{z}\right] \\ 
\left[ b_{x}\right] & \left[ b_{y}\right] & \left[ b_{z}\right] \\ 
\left[ c_{x}\right] & \left[ c_{y}\right] & \left[ c_{z}\right]%
\end{array}%
\]

\item Then the lattice vectors $\vec{u},\vec{v},\vec{w}$ are listed, one per
line, expressed in the coordinate system just defined:%
\[
\begin{array}{ccc}
\left[ u_{a}\right] & \left[ u_{b}\right] & \left[ u_{c}\right] \\ 
\left[ v_{a}\right] & \left[ v_{b}\right] & \left[ v_{c}\right] \\ 
\left[ w_{a}\right] & \left[ w_{b}\right] & \left[ w_{c}\right]%
\end{array}%
\]

\item Finally, the position $x_{i}$ and type(s) $t_{1i},t_{2i},\ldots $ of
atom for site $i$ are given, expressed in the same coordinate system as the
lattice vectors:%
\[
\begin{array}{cccc}
\left[ x_{a1}\right] & \left[ x_{b1}\right] & \left[ x_{c1}\right] & \left[
t_{11},t_{21}\ldots \right] \\ 
\left[ x_{a2}\right] & \left[ x_{b2}\right] & \left[ x_{c2}\right] & \left[
t_{12},t_{22},\ldots \right] \\ 
\vdots & \vdots & \vdots & \vdots%
\end{array}%
\]
\end{enumerate}

An example of such file, for an Al-Ti alloy adopting the hcp crystal
structure is:

\begin{center}
\begin{tabular}{ll}
\hline
\texttt{3.1 3.1 5.062 90 90 120} & (Coordinate system: $a$ $b$ $c$ $\alpha $ 
$\beta $ $\gamma $ notation) \\ 
\texttt{1 0 0} & (Primitive unit cell: one vector per line \\ 
\texttt{0 1 0} & expressed in multiples of the above coordinate \\ 
\texttt{0 0 1} & system vectors) \\ 
\texttt{0 0 0 Al,Ti} & (Atoms in the lattice) \\ 
\texttt{0.6666666 0.3333333 0.5 Al,Ti} &  \\ \hline
\end{tabular}
\end{center}

As an alternative to providing the above structural information, the user
can provide generators of the point group (which could also be the whole
point group) and the code will complete the full point group automatically.
This input file is called \texttt{sym.in} and has the format:

\begin{center}
\begin{tabular}{l}
\lbrack Number symmetry operations given as generators] \\ 
\lbrack 3$\times $3 matrix representing a symmetry operation in Cartesian
coordinates] \\ 
\emph{0 0 0} \\ 
\lbrack another 3$\times $3 matrix representing a symmetry operation in
Cartesian coordinates] \\ 
\emph{0 0 0} \\ 
\ldots%
\end{tabular}
\end{center}

\subsection{Output file}

The harmonics are output in the file \texttt{harm.out}, which has the
following format:

For each harmonic:

\begin{center}
\begin{tabular}{l}
\emph{1} \\ 
\emph{0} \\ 
\emph{0} \\ 
\emph{tensor} \\ 
\lbrack rank] \\ 
\emph{[3's repeated 'rank' times]} \\ 
\emph{[3}$^{\text{\emph{rank}}}$\emph{]} \\ 
\lbrack the 3$^{\text{rank}}$ elements of the tensor] \\ 
\ldots%
\end{tabular}
\end{center}

In addition, the standard output displays the harmonics in human-readable
format, with $x$, $y$, $z$ denoting the components of a unit vector.

\subsection{Command line options}

\begin{tabular}{ll}
-l=[string] & Input file defining the lattice (default: lat.in) \\ 
-s & Specify point group directly via generator (in sym.in) \\ 
-z=[real] & Tolerance for finding symmetry operations (default: 2e-4) \\ 
-r=[int] & Maximum Rank of the harmonic \\ 
-mr=[int] & Minimum Rank of the harmonic (default 1) \\ 
-sig=[int] & Number of significant digits printed (default: 5)%
\end{tabular}

Invoking the code without any options displays help. At the minimum, the
user must specify the -r option. All other options are optional.

\bibliography{avdw,avdwonly}

\end{document}